\documentclass[14pt]{article}
\usepackage{amsmath,amssymb,graphicx,authblk}
\usepackage[a4paper,margin=2cm]{geometry}

\usepackage{setspace}
\usepackage{lmodern}

\renewcommand{\vec}{\textbf}
\title{Twist-Induced Beam Steering and Blazing Effects in Photonic Crystal Devices}

\author[1,2]{Roy Nicolas}
\author[3]{Beicheng Lou}
\author[3]{Shanhui Fan}
\author[1,2]{Alexandre Mayer}
\author[2,4]{Michaël Lobet}
\affil[1]{Namur Institute for Complex Systems, University of Namur, Namur, Belgium}
\affil[2]{Namur Institute for Structured Matter, University of Namur, Namur, Belgium}
\affil[3]{Department of Applied Physics and Ginzton Laboratory, Stanford University, Stanford, California 94305, United States}
\affil[4]{John A. Paulson School of Engineering and Applied Sciences, Harvard University, 9 Oxford Street, MA-02138 Cambridge, USA}
\begin{document}
\maketitle
\begin{abstract}
Twisted bilayer photonic crystals introduce a twist between two stacked photonic crystal slabs, enabling strong modulation of their electromagnetic properties. The change in the twist angle strongly influences the resonant frequencies and available propagating diffraction orders with applications including sensing, lasing, slow light or wavefront engineering. 
In this work, we design and analyze twisted bilayer crystals capable of steering light in a direction controlled by the twist angle.
In order to achieve beam steering, the device efficiently routes input power into a single, twist-dependent, transmitted diffraction order. The outgoing light then follows the orientation of this diffraction order, externally controlled by the twist angle.
The optimization is performed using high-efficiency heuristic optimization method which enabled a data-oriented approach to further understand the design operation. The optimized device demonstrates an efficiency above $90\%$ across twist angles ranging from $0$ to $30$ degrees for both TE and TM polarizations. Extending the optimization to include left- and right-handed polarizations yields overall accuracy nearing $90\%$ when averaged across the entire 0 to 60 degrees control range.
Finally, we show how the device resembles blazed gratings by effectively canceling the undesired diffraction orders. The optimized devices exhibit a shared slant dependent on the selected diffraction order. Our analysis is supported by a structural blazing model arising from the data-oriented statistical analysis.
\end{abstract}


Twisted bilayer photonic crystals have garnered increasing attention due to their ability to introduce twist-dependent resonant frequencies, symmetry breaking, and control over their rich reciprocal lattice extended by the Moiré wavevectors \cite{zhang2023tphcbic,tang2023probetwistedband}. 
Associated remarkable phenomena include optical flat bands \cite{tang2021twphc,nguyen20221Dflatbands}, energy localization \cite{zadkov2023prl,wang2020localizationmoire}, optical singularities \cite{ni2024singul}, and lasing \cite{mao2021magicmoirelaser,luan2023moirelaser,raun2023magiclasergan}.
 The bilayer implementation enables on-chip reconfiguration of the twist for various optical applications, including beam shaping \cite{bernet14lens}, sensing \cite{tang2023onchip,zhang2018torsion}, tunable circular dichroism \cite{wu2017moire}, and frequency filtering \cite{lou2022tunableff}.  Additionally, microelectromechanical systems (MEMs) offer an effective solution for this reconfiguration \cite{tang2024onchip2D, arbabi2018mems}.

Beam steering refers to the control of the direction of a light beam. 
In photonic systems, beam steering can be achieved by active metasurfaces \cite{fourou2019bsdiel}, optical phased arrays \cite{ashraf2023metabeamsteer, lin2022beamsteer} or cascaded dielectric bilayers \cite{zhang2023bsms}. 
Recently, twisted bilayer photonic crystals were shown to be a promising approach by achieving efficient beam steering with a pair of dielectric layers only a few wavelengths thick \cite{lou24bs}.
Indeed, by adjusting the dielectric distribution within the layers, one can engineer diffraction efficiency \cite{gratingsbook}. 
While widely used blazed gratings operate on this principle, they lack tunability \cite{gao2021blazed}. 
In contrast, the proposed twisted bilayer photonic crystal device routes input power into a single twist-dependent diffraction order, resulting in a reconfigurable output beam. 
Here, the beam direction is directly controlled by the twist angle between the two layers.
By emitting in a single diffraction order, the device also inherits advantages of blazed gratings: 
they offer better beam control than phased arrays that actively manipulate the phase of multiple emitters. 
These arrays of emitters with a typical pitch larger than the wavelength, introduce multiple diffraction orders. 
Furthermore, the twisted device’s active surface is theoretically boundless and does not emit light itself \cite{gu2011bsslow}, allowing for seamless integration into on-chip systems.  
However, engineering such dielectric photonic crystals is a complex challenge. 
The twist angle adds an extra dimension of complexity to the design because the device lattice itself can change during operation. 
The freeform inverse design approach used in \cite{lou24bs} has led to intricate but high-performing designs, although their operational framework is not yet fully understood. 
Is the device's performance primarily attributable to the complex dielectric engineering, or is there a more fundamental operational principle at play? Addressing this question is crucial for advancing the development of any beam steering device based on twisted bilayer photonic crystals.

This work  aims to address the aforementioned question. Therefore, we provide herein a thorough theoretical analysis augmented by high-performance optimizations to design efficient beam steering devices.
Three templates of devices are explored: mini-layers \cite{lou24bs} and (in)homogenous combination of dielectric ellipses.
Our analysis is grounded in data-driven methodologies, leveraging high-throughput simulations \cite{cersonsky2021databip} to explore the design space effectively. 
Guided by particle swarm optimization (PSO) \cite{eiben2015introduction, eberhart1995particle, roy2024fuzzy}, we reveal that these optimized designs exhibit a consistent blaze angle tailored to the selected diffraction order.
Notably, our findings demonstrate that optimal devices mostly rely on a blazed configuration to suppress undesired diffraction orders. This point is further supported by an analytical model of the device.
These insights pave the way for advancing the design and fabrication of beam-steering twisted bilayer photonic crystals which could capitalize on established expertise in blazed gratings.
\section{Twisted gratings geometry}
\begin{figure}
  \centering
  \includegraphics[width=0.98\textwidth]{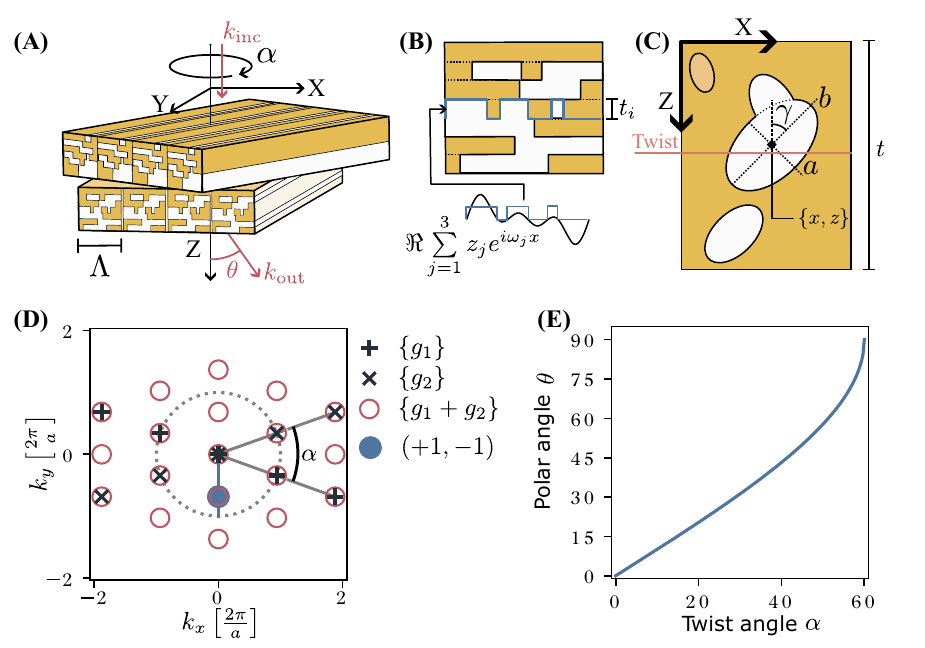}
  \caption{(A) Illustration of the twisted bilayer photonic crystal used for beam steering. 
  (B,C) Gratings and ellipses geometrical parameterizations are illustrated. 
  (D) The diffraction orders of both layers and those resulting from the twisted bilayer structure with an emphasis (blue dot and trajectory) on the targeted order in this work. 
  (E) The polar angle $\theta$ of the targeted order wavevector varying with a range of the control i.e. twist angle $\alpha$.}
  \label{fig:intro}
\end{figure}

The device, depicted in fig. \ref{fig:intro}(A), comprises two misaligned 1D photonic crystal slabs stacked in free space. 
Each slab is patterned with a unit cell of pitch $\Lambda$ in the $xz$-plane, following different templates illustrated in fig. \ref{fig:intro} (B, C).
These unit cell patterns are then extruded along the $y$-axis to obtain the layers. 
The device is designed to operate around a target wavelength, $\lambda$. 
The two twisted crystals possess the following reciprocal vectors sets
$$
\left\{\bf{g}_1\right\} = \left\{R(\alpha_1)i\frac{2\pi}{\Lambda}\vec x, \forall i \in \mathbb{Z}\right\}~\text{and}~\left\{\bf{g}_2\right\}=\left\{R(\alpha_2) j\frac{2\pi}{\Lambda}\vec x,\forall j \in \mathbb{Z}\right\}
$$
with $R(\alpha)$ a rotation matrix and $\vec x$ a unit vector along the $x$ direction. Here, we consider a twist $\alpha$ between the layers by taking  $\alpha_1=-\frac{\alpha}{2}$ and $\alpha_2=\frac{\alpha}{2}$.
The resulting twisted bilayer forms a hexagonal unit cell, with its truncated reciprocal space illustrated in fig. \ref{fig:intro}(D) as red circles. 
More generally, the set of reciprocal space vectors is
$$
\left\{\bf{g}_1+\bf{g}_2\right\} = \left\{(i+j)\cos\left(\frac{\alpha}{2}\right)\frac{2\pi}{\Lambda}\vec x+(j-i)\sin\left(\frac{\alpha}{2}\right)\frac{2\pi}{\Lambda}\vec y,\forall i,j \in \mathbb{Z}^2\right\},
$$
with $\vec x$ and $\vec y$ the unit vectors along $x$ and $y$ directions.
Below we denote each element in the set, which corresponds to a diffraction order, as $(i,j)$.
The design of the unit cell focuses on concentrating diffraction efficiency into a single Moiré diffraction order, $(+1,-1)$, as represented by the blue dot in fig. \ref{fig:intro}(D).
The targeted order could be both $(+1,-1)$ or $(-1,+1)$ as these exhibit twist-dependent magnitudes.
The magnitude of each reciprocal lattice  vector $\vec g$ directly affects the associated out-of-plane wavevector component
\begin{equation}
k_z = \sqrt{\left(\frac{2\pi}{\lambda}\right)^2-g_x^2-g_y^2},
\label{eq:kz}
\end{equation}
influencing light propagation.
Orders like $(0,0)$, $(0,1)$ and $(1,0)$ are unsuitable because their magnitudes are either zero or invariant with respect to $\alpha$.
Altough higher diffraction orders can exhibit similar steering capabilities, we focus on lower orders because a suitable choice of $\lambda$ relative to $\Lambda$ renders higher orders evanescent, as implied by eq. \ref{eq:kz}. 
Such a situation is met when $k_z$ is a complex number, which arises when
\begin{equation}
  \Lambda\lesssim\lambda.
\end{equation}
This strongly reduces the number of available orders in both reflection and transmission media,
keeping only three diffraction orders $(0,0)$, $(-1,1)$ and $(1,-1)$ in the light cone of fig. \ref{fig:intro}(D).
Tunable steering in $(+1,-1)$ is achieved as the outgoing beam inherits the polar angle dependency on the twist angle shown in fig. \ref{fig:intro}(E) and expressed as follows:
\begin{equation}
  \theta(\alpha)=\arcsin\left(\frac{\lambda}{\Lambda}2\sin{\left(\frac{\alpha}{2}\right)}\right).
  \label{eq:2dgrating}
\end{equation}
In practice, we choose a value at the threshold, $\lambda=1.01\Lambda$, since a higher value would make order $(+1,-1)$ evanescent for large twist angles. 
Here, the maximum twist $\alpha$ is $60$ degrees, as seen in fig. \ref{fig:intro}(D), any higher $\alpha$ would see the order become evanescent.
This is why fig. \ref{fig:intro}(E) stops just before $60^\circ$.

The designs under study are divided into three distinct templates. 
The first template, illustrated in fig. \ref{fig:intro}(B) treats the design as a stack of $N$ \emph{mini-layers} similar to \cite{lou24bs}. 
Each mini-layer's dielectric profile is parameterized by three complex phasors $(z_1,z_2,z_3)$ defining the grating in Fourier space (see SI for further details). 
The resulting signal is converted to a grating using thresholding with $\varepsilon_\text{low}=2$ and $\varepsilon_\text{high}=4$. 
This representation leads to $6$ free parameters plus an additional parameter for the mini-layer depth. A mini-layer parameterization thus resulting in $7N$ parameters.
The second and third templates consist of a set of $N$ ellipses arranged in the $xz$ plane as illustrated in fig. \ref{fig:intro}(C).  In the second template, each elliptical inclusion has a dielectric constant of $\varepsilon=4$, while in the third template, inclusions can take discrete $\varepsilon$ values from $1$ to $4$ . In both cases, the background permittivity is $\varepsilon_2=2$. Each ellipse is characterized by its position, axes length and slant, resulting in $5N$ free parameters. Additionally, the  layer thickness is a parameter, bringing the total to $5N+1$ parameters for the second template and $6N+1$  parameters for the third. Overlapping ellipses are combined using an \emph{OR} operation.

\section{Numerical approach}
The numerical setup that produces optimal designs is composed of two parts. First, extended rigorous coupled wave analysis (RCWA) is used to solve Maxwell's equations for the designs \cite{lou21rcwa}. Unlike classic RCWA, which requires twist angles that form a commensurate superlattice, this extended RCWA allows for the numerical characterization of twisted bilayers at any twist angle. By avoiding the supercell approach, this extension enables efficient (fast) numerical experiments by selectively computing a subset of scattering matrix coefficients. 
Second, an optimizer explores candidate designs in the parameter space. In particular, surrogate-assisted particle swarm optimization (PSO) is employed \cite{eberhart1995particle,roy2024fuzzy}. 
The surrogate model significantly reduces the number of RCWA simulations needed by leveraging faster predictions from a neural network model, as in \cite{wiecha21deepphot, roy23agpm}. Particle swarm optimization is particularly suited to continuous geometric parameters, but this implementation also handles categorical parameters. For example, it allows for material parameterization in the third template.

Particle swarm optimization requires the definition of a figure of merit for the diffraction efficiency in order to sort the designs from worst to best.
The performance of each design $\vec D$ is evaluated based on the transmission value  $T_{ij}$ of the targeted diffraction order (diffraction efficiency), averaged over twist angles from 0 to 60 degrees, with $N=50$ samples taken for the averaging. The figure of merit is thus a function of the design, diffraction order $(i,j)$ and twist angle
$$
f(\vec D) = \frac{1}{N}\sum_{k=0}^{N-1} T_{ij}(\alpha_k; \vec D),~\text{with}~\alpha_k=k\frac{60^\circ}{N-1}.
\label{eq:fom}
$$
While we focus on the $(i,j)=(+1,-1)$ order, we will show that all conclusions can be adapted to the $(-1,+1)$ order through symmetry.
The combination of extended RCWA with PSO facilitates rapid computation of this figure of merit (2 seconds) in highly parallel workloads (16 threads). Therefore, 1000 devices were designed for each of the three different parameterization templates. In total, $4.5$ million devices were evaluated, among which 450 thousand were simulated.

\begin{figure}

  \includegraphics[width=0.98\textwidth]{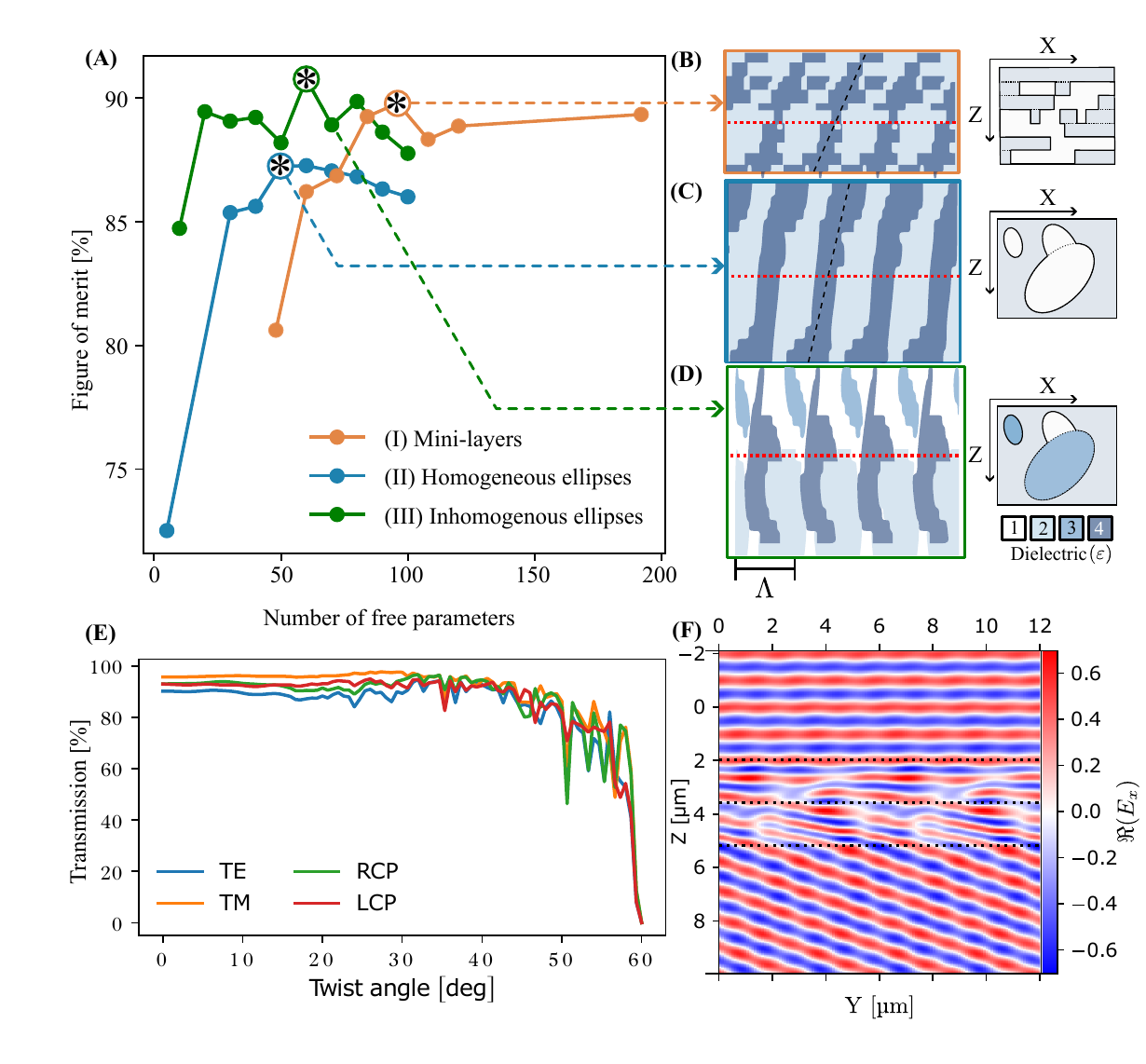}
  \caption{(A) Performance of the different templates considered as a function of the number of free parameters. (B-D) Best devices obtained for the mini-layer, ellipses and heterogenous ellipses templates respectively. The plane separating the two twisted photonic crystals is drawn in red. (E) The transmission in order $(+1,-1)$ for all twist angle values of the best design in the homogenous ellipse template (C). (F) The $\Re E_y$ field map for the previous design with \textit{Y} polarization.}
\label{fig:perfdata}
\end{figure}

Figure \ref{fig:perfdata}(A) shows the optimal figure of merit $f^*$ for each template relative to the number of free parameters, while fig. \ref{fig:perfdata}(B, C, D) display the dielectric profiles for the best configurations of the three design templates. 
The optimal figure of merit $f^*$ is averaged across 4 polarizations (\textit{X},\textit{Y},\textit{RCP} and \textit{LCP}). 
The mini-layers template reaches $f^*=90\%$, slightly outperforming previous work ($88\%$) that optimized the \textit{X} and \textit{Y} transmission \cite{lou24bs}. 
The ellipses designs provide simpler, more continuous structures yet still efficient, peaking just above $87\%$. 
This template was motivated by the tendency of the mini-layers template to produce structures roughly continuous along the $z$ axis, even across the twist plane.
When allowing the optimizer to choose between a discrete set of material, the efficiency peaks at $91\%$, due to the greater freedom. 
With a fabrication procedure in mind, the parameterization can be constrained to match feasibility constraints, i.e., choosing from a pool of available materials for each component.
Regardless of the template, the devices primarily consist of a roughly continuous slanted dielectric structure, tapering to thinner edges near the limits of the layers.
When present, these thinner edges are improving the impendance matching with air, facilitating in- and outcoupling of light.
While the optimal design for homogenous ellipses does not have this feature, it illustrates very well the slanted and continuous dielectric profile along $z$.

As the number of free parameters increases, we initially observe an improvement of the performance. 
The performance however saturates and even decreases at some point for all templates due to two factors: 
the limited performance gain from a finer design and the intrinsic limitation of the optimizer to find the best solutions in an increasingly larger space.
This limitation is present in all optimizers, known as the \emph{curse of dimensionality}\cite{chen2015measuring}.

As previously stated, fig. \ref{fig:perfdata} reveals a shared slant angle for template I and II, indicated by a dashed black line, confirmed by analysis of all optimized devices (see SI).
This slant angle ranges from $\gamma_{\text{num}}=15^\circ$ to $24^\circ$. For the alternate diffraction order $(-1,+1)$, the optimal slant angle reverses, suggesting a fundamental relationship between slant angle and diffraction order.
The following section justifies the existence of this slant angle using a structural blazing model.

\section{Reduced model of the slanted geometry}

The previous section highlights how various optimizations lead to shared features in the optimal designs. 
Specifically, devices exhibit a common slant angle that is dependent on the targeted diffraction order. 
From diffraction theory, the diffraction orders when going from a medium of index $n_{i}$ to $n_{i+1}$ are function of the periodicity of the grating, 
the wavelength and the incidence angle ($\theta_i$). 
This dependence is expressed by the gratings equation \cite{born_wolf_1999}
\begin{gather}
n_{i+1}\sin\theta_{i+1,m}=n_{i}\sin\theta_i-m\frac{\lambda}{\Lambda},\\
\text{with } m \in \mathbb{Z}.\notag
  \label{eq:gratings}
\end{gather}
While this equation predicts diffraction orders at discrete angles $\theta_{i+1,m}$, no clue is given of their intensity. 
Although specialized theories exist for the diffraction efficiency of common grating shapes \cite{gratingsbook,gratingsbook2}, 
solving Maxwell's equations is required for the most general case. The structures studied in the present work fall under this general category, 
but the presence of a common slant angle makes them more closely resemble blazed gratings. 
Blazed gratings are gratings with a specific slanted profile that aim at concentrating the diffracted light in a specific diffraction order \cite{gao2021blazed}. 
In this section, we prove that the slanted geometry is at the core of the device operation by applying blazing principles to a simplified structural model.

\begin{figure}
  \centering
  \includegraphics[width=0.98\textwidth]{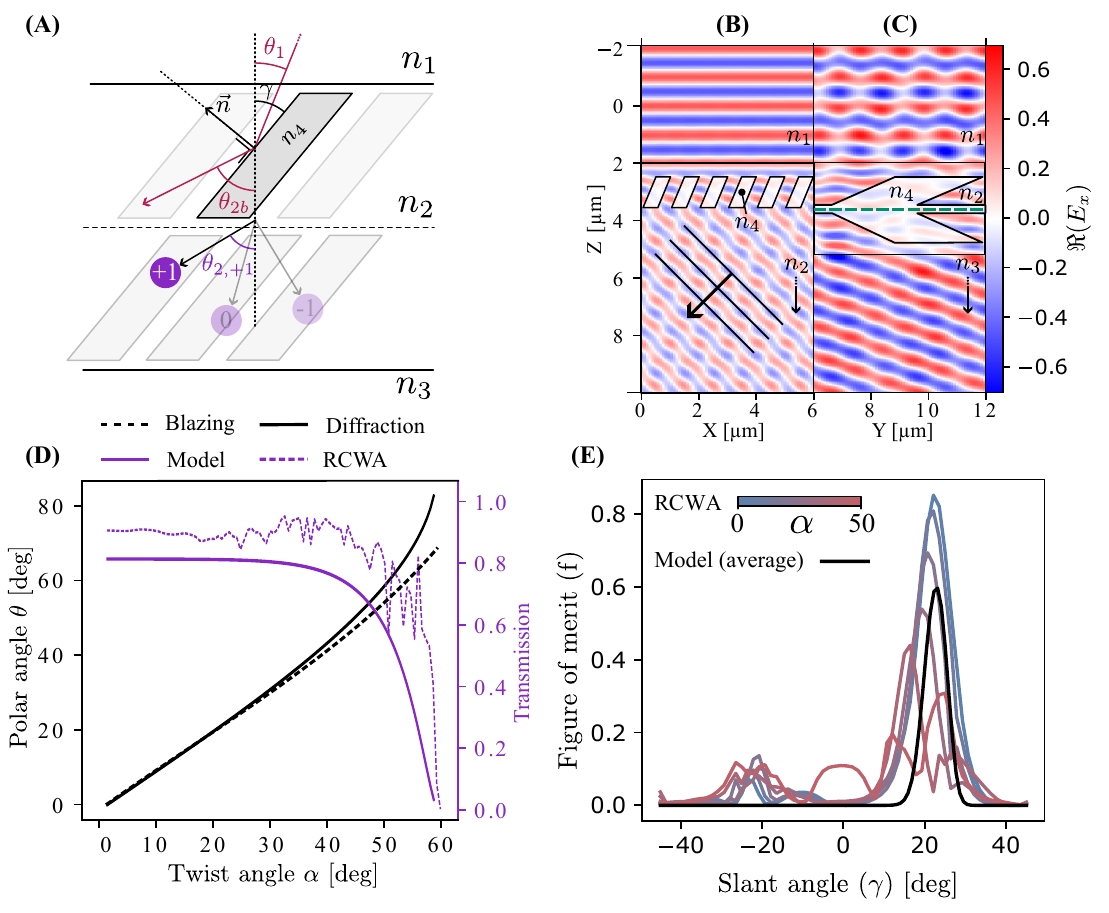}
\caption{
(A) Scheme of the reduced structural model. 
(B) Field map for \textit{Y} polarization with only one layer on a substrate of index $n_2$.
(C) Field map for \textit{Y} polarization with both layers twisted by an angle $\alpha=20^\circ$ in air. 
  (D) The diffraction angle from the grating equation (\ref{eq:gratings}) compared to the blazing angle formula (\ref{eq:finalblaze}). On a second axis, the diffraction efficiency is shown for our reduced model and RCWA simulation. 
(E) The figure of merit for different slant angles computed using our model compared to RCWA simulations for the different twist angles.}
\label{fig:reduced}
\end{figure}

The structural model simplifies the $xz$ profile of the unit cell used in each layer in order to isolate the blazing effect as illustrated in fig. \ref{fig:reduced}(A).
Each unit cell consists in a parallelogram of index $n_4=2$ on a background $n_2=\sqrt{2}$. 
The model discussed is valid for a wide range of parallelogram widths ($0.2<\frac{w}{\Lambda}<0.8$) and heights ($h>\Lambda$). These dimensions have minimal impact on the subsequent analysis as long as the layers can be considered diffraction gratings.
An illustration of their influence is available in SI. When comparing our model to RCWA in fig. \ref{fig:reduced}(B,C,D), the height is fixed to $4.0\Lambda$ and the width to $0.4\Lambda$. Again, without impacting the phenomenon discussed, we can consider the presence of a buffer interlayer of index $n_2=\sqrt{2}$ with a depth of $0.2\Lambda$ between the two layers.
The refractive indices in emergence and incidence media are kept the same ($n_1=n_3=1$), 
so that diffraction orders for the structural model can be analyzed by considering the same two successive diffraction events as for the optimized structures.
The diffraction polar angles are still governed by eq. \ref{eq:2dgrating}, which is charted in fig. \ref{fig:reduced}(D).

In the following developpement, we will treat the diffraction through each layer separately.
The incoming beam is considered at normal incidence such that the incoming polar angle is $\theta_1=0^\circ$,
while $\theta_2$ is the orientation after diffraction through the first layer and finally $\theta_3$,
the orientation in free space after the crystal which is given by eq. \ref{eq:2dgrating}.
Starting with the upper grating, eq. \ref{eq:gratings} indicates three propagative transmitted diffraction orders in the buffer layer $n_2\approx\sqrt{2}$, each with distinct emergence angles
\begin{gather}
\theta_{2,m}=\arcsin{(-m\frac{\lambda}{n_2\Lambda})}\approx-m~45^\circ \\
\text{ with } m\in \left\{-1,0,1\right\}.\notag
\label{eq:difforders1}
\end{gather}
We now analyze the power distribution among the available diffraction orders using a blazing heuristic.
Blazing occurs in a specific order when the diffraction angle coincides with the specular reflection against the grating's slanted surface. 
Similarly, blazing occurs with refraction through the same surface. 
From empirical analysis, blazing in the system occurs by reflection.
This phenomenon is best illustrated by the field maps of fig. \ref{fig:reduced}(B) where order $m=+1$ is seen blazing against grating fins oriented at $\gamma=22.5$ degrees for an outgoing wavevector at $\theta_{2b}=-45^\circ$.
This field map is obtained by RCWA, deposing the upper, untwisted layer on a substrate of index $n_2$ to isolate its diffraction properties. 
The observed outgoing polar angle closely aligns with the $(+1)$ diffracted angle of eq. \ref{eq:gratings}, where $\theta_{2,+1}=-45^\circ.$
Analytically, the blaze angle is derived from specular reflection against the surface slanted with an angle $\gamma$ of the parallelogram illustrated in fig. \ref{fig:reduced}(A)
\begin{equation}
\theta_{2b}(\gamma) = \theta_1 - 2\gamma
  \label{eq:raytracing1D}
\end{equation}
For blazing to occur, the blazing slant angle $\gamma_b$ is tuned to align the blazing angle ($\theta_{2b}$) with order $+1$
\begin{equation}
\theta_{2b}(\gamma_b) = \theta_{2,+1} = -45^\circ.
\end{equation}
This relation is satisfied for a specific slant $\gamma_b=22.5^\circ$. A value that aligns with the range of $\gamma_\text{num}=15-24^\circ$ from the previous numerical study.

The second grating, without being twisted ($\alpha=0$), has an identical blazing configuration to the first one.
When describing the second layer, we consider that the first layer blazing is perfect: the output angle can be assimilated to the diffraction angle ($\theta_2=\theta_{2b}=\theta_{2,+1}$).
As the two layers share a common slant $\gamma$ the blaze angle $\theta_{3b}$ is expected at $\theta_{2}-2\gamma=0^\circ$. 
Consistently, the diffraction angle for this second diffraction event is obtained from eq. \ref{eq:gratings} with an incidence at $\theta_2=45^\circ$ relative to the grating normal.
At this angle, only order $(-1)$ propagates, with $\theta_{3,-1}=0^\circ$, equal to the blaze angle $\theta_{3b}$.
The beam direction is therefore unchanged when traversing the untwisted bilayer through order $(+1,-1)$.

For a nonzero twist angle $\alpha$, the model requires to consider the three dimensions of space
as the orientation of the slanted surface of the grating depends on the twist angle $\alpha$.
This orientation is defined by the surface normal $\vec n(\alpha)$.
The normalized outgoing wavevector $\vec u_r$ is obtained from the incident wavevector $\vec u$ on the second grating using a vectorial form of eq. \ref{eq:raytracing1D} commonly used in ray tracing\cite{pharr2016pbr}
$$
\vec u_r=\vec u - 2\left[\vec n(\alpha) \cdot \vec u \right]\vec n(\alpha).
$$
The outgoing polar angle $\theta_{3b}$ is obtained by evaluating the angle between $\vec u_r$ and the layer normal $\vec e_z$ while accounting for refraction
\begin{equation}
\theta_{3b}= \arcsin\left(\frac{n_2}{n_3}\sqrt{1-\left(\vec u_r\cdot \vec e_z\right)^2}\right).
\label{eq:finalblaze}
\end{equation}
This blazing angle is compared to the diffraction angle in fig. \ref{fig:reduced} (D) for a slant angle $\gamma=22.5^\circ$. 
For small twist angles, the relation is mostly linear (below $30^\circ$), the match between blaze and diffraction angles is excellent.
For twist angles superior to $30^\circ$, the blaze and diffraction angles start to diverge.
While this first formula indicates that the blazing will be less efficient for large twist angles, we do not yet have a clear picture on the impact of this divergence.

The impact of misalignment on transmission values is modeled using a gaussian distribution centered around the blaze angle:
\begin{equation}
  t_i(\theta)=Ae^{-\frac{\left(\theta-\theta_{ib}\right)^2}{2\sigma^2}}.
  \label{eq:gaussian}
\end{equation}
Here, the parameters $A$ and $\sigma$ are derived by fitting RCWA simulations of a single layer like the one on fig. \ref{fig:reduced}(B). 
The standard deviation ($\sigma\approx17^\circ$) quantifies the sensitivity of transmission to misalignment between blaze and diffraction angles. The amplitude $A$ approximates the transmission coefficient for a slant angle in ideal alignment $(\Delta \theta=0)$. This model assumes transmission approaches zero when the blaze and diffraction angles are significantly misaligned.
The resulting transmission is given by
\begin{equation}
t = t_1(\theta_{2b}-\theta_2)t_2(\theta_{3b}-\theta_3),
\end{equation}
which accounts for transmission losses due to misalignment in both layers. 
This approximation is compared  in fig. \ref{fig:reduced}(E) to RCWA simulation for the optimal device in the ellipses template across varying twist angles. 
Our model respects the profiles observed during optimization and successfully predicts the steep downward trend between $45$ and $60$ degrees.
The optimized device used for comparison improves upon the reduced model over the whole control range.
Compared to our reduced model, optimized designs tend to use curved surfaces along $z$, unlike our parallelograms, offering a distribution of slant angles $\gamma(z)$.
Altough it is not the case for the best homoegenous ellipses design, designs exhibit thinner higher dielectric regions as we get close to the outer surface of the layers.
This phenomenon is due to impendence matching, much like in light extraction optimizations in light-emitting diodes inspired by fireflies\cite{bay2013optimal,mayer2015galed}.
It is noteworthy that the blazing behavior alone in this simple structure yields a figure of merit around $70\%$.
The field map in fig. \ref{fig:reduced}(C) shows the reduced device in operation at $\alpha=20^\circ$, with, indeed, a convicing plane wave going out of the device.
The optimizer then further enhances this blazing effect, even for a single homogenous ellipse, leading to $73\%$ efficiency.
This result is coherent as the improvement, for a relatively similar device compelexity, comes mainly from the fine tuning of the orientation and size of the dielectric inclusion.

The model's figure of merit—defined in eq. \ref{eq:fom} as the average transmission over the full range of twist angles—is plotted for various slant angles in fig. \ref{fig:reduced}(E).
The results show a clear optimum at $\gamma\approx23^\circ$, consistent with previous numerical predictions. 
As a benchmark, RCWA-computed figure of merit for this reduced model is also shown for a set of twist angles.
Interestingly, it confirms that this blazing behavior applies uniformly across twist angles ranging from $0$ to $30$ degrees. 
The optimal slant $\gamma$ seems to decrease with increasing twist $\alpha$, supporting the hypothesis that a distribution $\gamma(z)$ is indeed benefitial to optimized designs.
This stability of the blazing behavior is key to the device operation as we already get a figure of merit around $65\%$ for this reduced model.
The model does not capture the smaller RCWA peaks in fig. \ref{fig:reduced}(E), around $0^\circ$ and $\gamma\approx-20^\circ$, mainly due to the assumptions in eq. \ref{eq:gaussian}.
Indeed, we neglect blazing through and against the other surfaces of the parallelogram as well as blazing that also happens on the other (incidence) side of the crystal.
However, these phenomena are of lesser magnitude, allowing our model to successfully explain the behaviour of the crystal.
We conclude that this structural model of the identified blazing behavior underscores the importance of the blazed configuration.

\section{Conclusions}
Firstly, our study demonstrates that heuristic optimization enables the rapid design of high-efficiency beam-steering devices based on twisted bilayers photonic crystals. 
Mini-layered structures achieved up to $90\%$ beam steering efficiency. 
Another, more organic template achieves a slightly lower efficiency ($87\%$), but offers a simpler design. 
Similarly, solutions based on discrete material choices offer enhanced ($91\%$) performance. 
Though slightly less manufacturable than the ellipses design, they still present promising prospects for practical fabrication.

Secondly, we found that the optimal devices rely on a bilayer blazing effect, where a specific slant angle allows sequential blazing into the $(+1)$ and then $(-1)$ orders of the first and second layers. The change in twist angle alters the reciprocal space geometry, adjusting beam orientation while preserving the blazing condition for twists up to 45 degrees. 
We proved using a reduced model that this preservation of the blazing condition is structural.
For small twist angles, the diffraction and blazing angles are predominantly linear and align perfectly.
The model also predicts the steep decrease in diffraction efficiency for large twist angles and identifies an optimal slant angle of approximately $23^\circ$.
Optimized devices produce diffraction efficiencies consistent with this model but achieve higher overall performance. These devices consistently feature slant angles close to the model’s predicted value. The performance enhancement appears to result from two key mechanisms: impedance matching through engineering of the refractive index distribution and variation of the structure's slant angle using curved geometries along the crystal's normal direction. 

\section{Data availability}

{\bf{Acknowledgements}} S. F. acknowledges the support of a MURI project from the U. S. Air Force Office of Scientific Resarch (Grant No. FA9550-21-1-0312). A.M. and M.L. are Research Associates of the Fonds de la Recherche Scientifique – FNRS. Computational resources have been provided by the Consortium des Équipements de Calcul Intensif (CÉCI), funded by the Fonds de la Recherche Scientifique de Belgique (F.R.S.-FNRS) under Grant No. 2.5020.11 and by the Walloon Region. The present research also benefited from computational resources made available on Lucia, the Tier-1 supercomputer of the Walloon Region, infrastructure funded by the Walloon Region under the grant agreement n°1910247.
\bibliographystyle{unsrt}
\bibliography{biblio}
\end{document}